\documentclass[pdflatex,sn-mathphys-num]{sn-jnl}

\usepackage{graphicx}%
\usepackage{multirow}%
\usepackage{amsmath,amssymb,amsfonts}%
\usepackage{amsthm}%
\usepackage{mathrsfs}%
\usepackage[title]{appendix}%
\usepackage{xcolor}%
\usepackage{textcomp}%
\usepackage{manyfoot}%
\usepackage{booktabs}%
\usepackage{algorithm}%
\usepackage{algorithmicx}%
\usepackage{algpseudocode}%
\usepackage{listings}%
\usepackage{lineno}

\theoremstyle{thmstyleone}%
\theoremstyle{thmstyletwo}%

\theoremstyle{thmstylethree}%

\begin{document}

\title[Article Title]{Perfecting Imperfect Physical Neural Networks with Transferable Robustness using Sharpness-Aware Training}


\author[1]{Tengji Xu}
\equalcont{These authors contributed equally to this work.}\email{tengjixu@link.cuhk.edu.hk}
\author[1]{Zeyu Luo}
\equalcont{These authors contributed equally to this work.}\email{zeyuluo@cuhk.edu.hk}
\author[1]{Shaojie Liu}
\equalcont{These authors contributed equally to this work.}\email{shaojieliu@link.cuhk.edu.hk}
\author[1]{Li Fan}
\equalcont{These authors contributed equally to this work.}\email{lifan210@link.cuhk.edu.hk}
\author[1]{Qiarong Xiao}\email{qrxiao@link.cuhk.edu.hk}
\author[1]{Benshan Wang}\email{bswang@link.cuhk.edu.hk}
\author[1]{Dongliang Wang}\email{dongliangwanh@link.cuhk.edu.hk}
\author*[1]{Chaoran Huang}\email{crhuang@ee.cuhk.edu.hk}

\affil*[1]{\orgdiv{Department of Electronic Engineering}, \orgname{The Chinese University of Hong Kong}, \orgaddress{\street{Shatin}, \city{Hong Kong SAR}, \country{China}}}


\abstract{AI models are essential in science and engineering, but recent advances are pushing the limits of traditional digital hardware. To address these limitations, physical neural networks (PNNs), which use physical substrates for computation, have gained increasing attention. However, developing effective training methods for PNNs remains a significant challenge. Current approaches, regardless of offline and online training, suffer from significant accuracy loss. Offline training is hindered by imprecise modeling, while online training yields device-specific models that can't be transferred to other devices due to manufacturing variances. Both methods face challenges from perturbations after deployment, such as thermal drift or alignment errors, which make trained models invalid and require retraining. Here, we address the challenges with both offline and online training through a novel technique called Sharpness-Aware Training (SAT), where we innovatively leverage the geometry of the loss landscape to tackle the problems in training physical systems. SAT enables accurate training using efficient backpropagation algorithms, even with imprecise models. PNNs trained by SAT offline even outperform those trained online, despite modeling and fabrication errors. SAT also overcomes online training limitations by enabling reliable transfer of models between devices. Finally, SAT is highly resilient to perturbations after deployment, allowing PNNs to continuously operate accurately under perturbations without retraining. We demonstrate SAT across three types of PNNs, showing it is universally applicable, regardless of whether the models are explicitly known. This work offers a transformative, efficient approach to training PNNs, addressing critical challenges in analog computing and enabling real-world deployment.}




\maketitle

\section{Introduction}\label{sec1}

The tremendous success of modern artificial intelligence (AI) is driven by a synergic operation of hardware and algorithms~\cite{lecun2015nature}.  Powerful hardware, particularly graphics processing units (GPUs), has revolutionized neural network (NN) computing by drastically reducing the time and resources needed to process large datasets. Central to this progress is backpropagation (BP), the foundational algorithm for training NNs, which efficiently dictates how NNs to learn and evolve by minimizing errors~\cite{rumelhart1986nature}. 

\begin{figure*}[ht]
\centering
\includegraphics[width=0.95\textwidth]{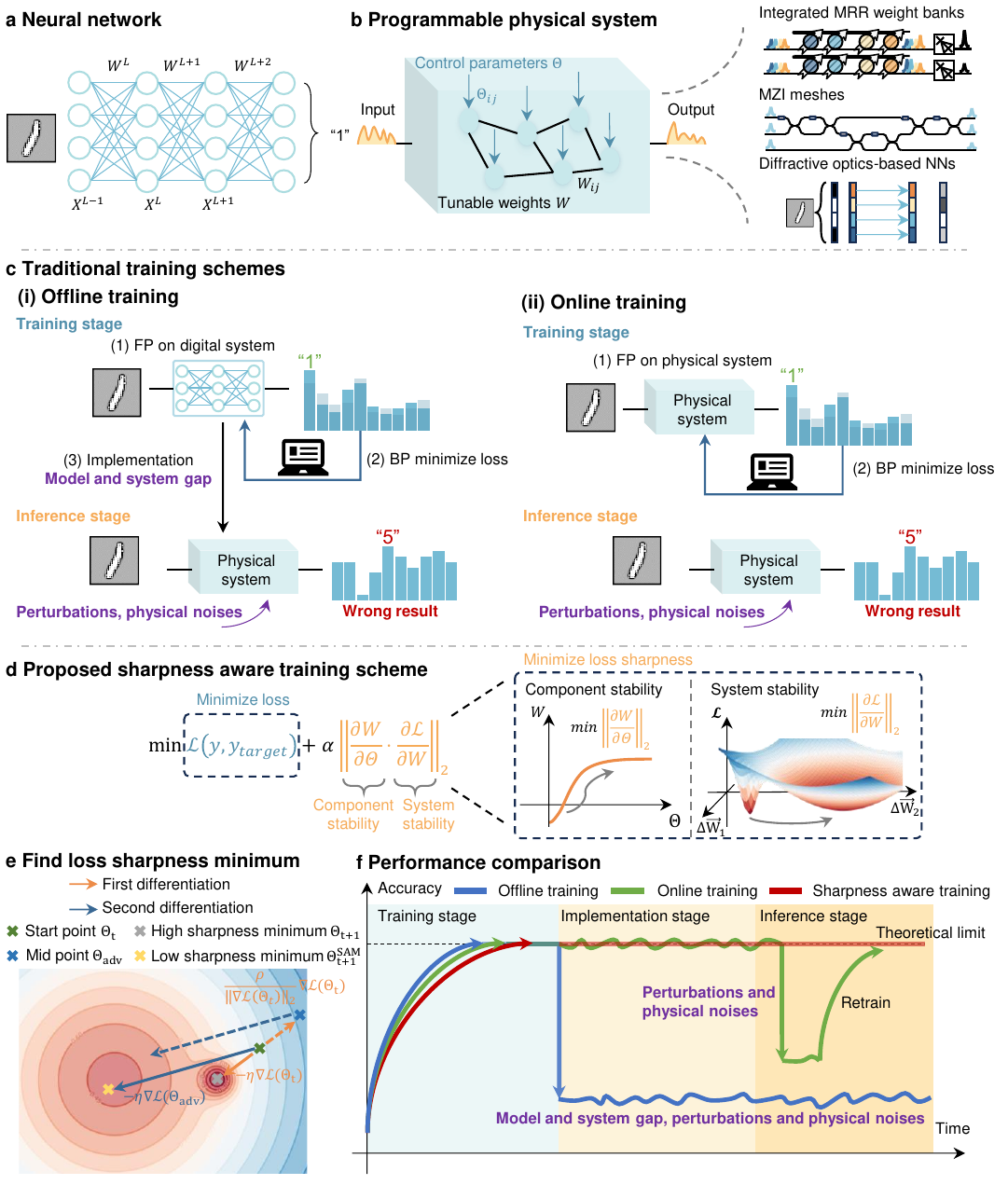}
\caption{The detailed comparison of physical neural network training methods and the proposed sharpness aware training method’s principle.  (a) Diagram illustrating a typical neural network. The neural network contains synapses to perform matrix-vector multiplications followed by nonlinear activation functions. (b) Schematic diagram of a programmable physical system with control parameters $\Theta$ and tunable weights $W$. The tunable weights $W$ are directly controlled by the control parameters $\Theta$. The system takes input signals and produces output signals based on the adjusted parameters. (c) Illustration of offline training and online training methods. The tunable weights are achieved by separately controlling different physical parameters including currents and phases. (d) Proposed Sharpness-Aware Training scheme. The training goals include reducing loss while increasing component and system stability, respectively. (e) Schematic diagram of the proposed sharpness aware training scheme parameter update. (f) Performance comparison between offline training, online training, and Sharpness-Aware training. MRR: Micro-ring resonator. MZI: Mach-Zehnder Interferometer. NN: Neural network. FP: Forward propagation. BP: Backpropagation.}
\label{fig1}
\end{figure*}

Recent AI models begin to strain the capabilities of traditional digital hardware. The cost of training and inferring AI models doubles every two months, significantly outpacing the Moore’s Law. Consequently, AI systems are increasingly facing challenges caused by hardware efficiency. To address hardware limitations, a new computing paradigm, neuromorphic computing has gained significant traction. Many neuromorphic computers are physical systems leveraging photonics, analog electronics, and other wave-based physics, as shown in Figure.\ref{fig1}a and Figure.\ref{fig1}b~\cite{momeni2023science,wright2022nature,wetzstein2020nature,romera2018nature,chen2020nature,grollier2020ne}. The physical models of these systems exhibit mathematical isomorphism with NNs, enabling their computing capabilities. Their physical configurations can alleviate the data movement bottleneck found in traditional digital hardware, substantially improving both computing speed and energy efficiency. Large-scale, high-speed, and energy-efficient physical neural networks (PNNs) have been widely reported~\cite{xu2024science,xia2024np,yildirim2024np,fu2024light,chen2023nature,youngblood2023np,chen2023np,ashtiani2022nature,zhou2022light,huang2021ne,feldmann2021nature,shastri2021np,zhou2021np,lin2018science,shen2017np}.

However, the development of suitable training methods that are synergistic with these new computing hardware has lagged behind. The training process remains cumbersome and unreliable for PNNs. Traditional gradient-based algorithms, such as BP, cannot be directly applied to PNNs due to the difficulty of obtaining gradients from physical systems. Moreover, even with appropriate training methods, PNNs struggle to maintain computing accuracy during deployment, as physical systems are inherently highly susceptible to disturbances compared to digital hardware, leading to errors~\cite{momeni2023science,vadlamani2023sa,xu2024optica}.

In reviewing the existing training methods, they can be divided into two main categories: offline training and online training (Figure.\ref{fig1}c). To train a PNN offline, the process begins by extracting a digital model of the physical system. The training is then conducted on a digital computer, typically using BP. Once the model is trained, the parameters are applied to the PNN. However, offline training requires highly accurate digital modeling of the physical system, which is challenging due to unavoidable fabrication-induced errors, and intrinsic noise present in the physical systems. As a result, the inference accuracy is often significantly lower than the theoretical value, as illustrated in Figure.\ref{fig1}c(i). In contrast, online training alleviates the need for precise digital modeling by looping the physical system directly into the training process. However, online training faces the problem of obtaining the gradients of the physical system. Current methods for obtaining gradients include: (1) Approximating the gradients using a digital twin model of the physical system obtained through data-driven methods~\cite{zheng2023nmi,wright2022nature,spall2022optica,zhan2024lpr,huo2023nc}; (2) Estimating the gradients through finite-difference methods by applying perturbations and measuring difference~\cite{cheng2024nc,wan2024oea,bandyopadhyay2022arxiv}; (3) Directly measuring the gradients physically~\cite{xue2024nature,pai2023science,wanjura2024nphysics,hughes2018optica,zhou2020pr,hermans2015nc}. Each of these approaches introduces overhead, either through excessive measurement requirements or the computational cost of creating a digital twin model. To mitigate this overhead, various gradient-free approaches have been proposed~\cite{momeni2023science,nakajima2022nc,filipovich2022optica}. However, their training efficiency and effectiveness in complex applications have yet to be fully demonstrated. Moreover, a significant drawback of all online learning approaches is that, because the specific physical system is looped into the training process, the trained results are only applicable to that particular system and cannot be transferred to other systems, even those with the same
design. Our simulation results show the online-trained model is even more vulnerable to imperfections than the offline-trained model when transferring the model to other systems. A critical issue across both offline and online training methods is that any perturbations after deployment, such as thermal drift or alignment errors, will make the trained parameters ineffective, leading to significant accuracy loss and requiring retraining~\cite{momeni2023science}. 

This work proposes and demonstrates a novel training technique that overcomes the challenges of both offline and online training. It enables accurate training of PNNs using efficient BP algorithms, without the need for an exact digital model or looping the actual hardware during the process. The systems trained offline by SAT surprisingly outperform those trained online, even in the presence of significant modeling and fabrication errors. Additionally, SAT addresses a key limitation of current online training methods. Once trained, the parameters can be reliably deployed to other devices for inference without accuracy loss, even if they exhibit slight differences due to fabrication variances. Equally important, the technique is highly resilient to perturbations after deployment, allowing PNNs to continue operating accurately under perturbations without the need for retraining.

The key to these benefits is that our training method leverages the geometry of the loss landscape in a physical system~\cite{li2018nips,foret2021iclr}. Our method minimizes not only the loss value but also the loss sharpness, as shown in Figure.\ref{fig1}d. By identifying parameters in regions of uniformly low loss, our approach ensures that the loss remains minimal despite challenges common in analog computing systems, such as modeling inaccuracies, fabrication variances, and external perturbations. We call our method Sharpness-Aware Training (SAT), inspired by the machine learning technique Sharpness-Aware Minimization~\cite{foret2021iclr}, which enhances the generalization capabilities of models. SAT is universally applicable to various PNNs, regardless of whether their models are explicitly known or unknown. We demonstrate its versatility in three typical optical neural networks (ONNs): integrated microring resonator (MRR) weight banks, Mach-Zehnder interferometer (MZI) meshes, and diffractive optics-based NNs. These demonstrations highlight the effectiveness and accuracy of SAT during both training and deployment stages, even under error sources inherent in analog computing. This includes imprecise modeling prior training, thermal and electrical disturbances that heavily affect sensitive devices such as resonators, fabrication variances in integrated optics, and deployment errors commonly seen in free-space optical systems. Overall, SAT offers a practical, effective, and computationally efficient solution for training and deploying PNNs in real-world applications.


\clearpage

\section{Results}\label{sec2}
\subsection{Sharpness-Aware Training: Seeking loss minimum and loss sharpness minimum in a physical system}

A deep neural network (DNN) is a multi-layer system, where each layer consists of synapses and nonlinear nodes, as depicted in Figure.\ref{fig1}a. The mathematical model of the $l$-th layer of DNN is given by,

\begin{equation}
X^{l+1} = A^{l}(W^{l}X^{l}+b^{l})
\label{eq1}
\end{equation}

Here $X^{l}$ is the input of the $l$-th layer, $W^{l}$ and $b^{l}$ denote the trainable weights and bias, and $A^{l}(\cdot)$ represents the $l$-th layer nonlinear activation layer. A PNN emulates this process through a physical system controlled by a set of tunable physical parameters $\Theta$, such as electrical currents. These parameters determine the system’s transfer function, which can be expressed by $y=F(x;\Theta)$  with $x$ and $y$ representing the system input and output and $\Theta$ representing the tunable physical parameters. Tuning these control parameters affects the system's transfer function, making them equivalent to changing weights in a DNN. Therefore, training a PNN is to find the optimal control parameters that minimize the loss function for a given task.

To train such a system accurately, obtaining the exact mathematical expression for $y=F(x;\Theta)$ is essential for conventional methods. However, this is nearly impossible for real-world physical systems, which are governed by highly complex physics and are prone to parameter drift, device crosstalk, and environmental disturbances. Discrepancies between the theoretical model and the actual system lead to training errors. Online training reduces the reliance on an exact mathematical expression. It obtains the true loss function by experimental measurement. With the true loss function, even if there are slight discrepancies between the hardware and the theoretical model, the loss can still be minimized through iterative optimization~\cite{wright2022nature}. However, online training is inefficient because it is specific to the particular system being trained and cannot be easily transferred to other devices, even those fabricated with the same design. This limitation is unfavorable compared to the typical functioning of modern AI hardware, where training is conducted on a single high-end computer or a cloud-based computer cluster, and the trained parameters are then deployed across many edge devices for inference~\cite{vadlamani2023sa}. Moreover, a critical issue across both offline and online training methods is that any perturbations after deployment, such as thermal drift or installment errors, can render the trained parameters invalid.

We address all the challenges of both offline and online training with our SAT method, which leverages the geometry of the loss landscape in a physical system to modify the training objective accordingly. The loss landscape, illustrated in Figure.\ref{fig1}e, represents how the loss changes as parameters are adjusted. Traditional training methods like BP only focus on minimizing the loss function, which typically ends at the ‘High sharpness minimum’ as depicted in Figure.\ref{fig1}e. The ‘High sharpness minimum’ is characterized by sharp curvatures, where small changes in parameters lead to rapid increases in loss. In machine learning, such suboptimal minima lead to poor model generalization. In this work, for the first time, we associate these sharp minima in the loss landscape with the poor robustness of physical systems. At such sharp minima, slight parameter changes due to modeling discrepancies, fabrication variances, or environmental disturbances can cause rapid increases in loss and errors, affecting both the training and deployment stages when using offline and online training methods.

Our SAT method addresses these problems by searching for flat minima in the loss landscape, as the ‘Low sharpness minimum’ illustrated in Figure.\ref{fig1}e. At these flat minima, the loss remains minimal even with significant parameter changes, providing high resilience and robustness against errors and imperfections in physical systems. In this work, we derive a universal framework for finding flat minima across different PNN systems. 

To find the ‘Low sharpness minimum’, the goal is not only to minimize the loss but also to ensure parameters lie in the neighborhoods with uniformly low loss value. The optimization process of SAT can be mathematically expressed as to find $\Theta$ minimizing the loss function $L_1$, as shown in Equation.\ref{eq2}:

\begin{equation}
L_{1} = L(y,y_{target};\Theta)+\alpha ||\frac{\partial L(y,y_{target};\Theta)}{\partial \Theta}||_{2}
\label{eq2}
\end{equation}

The first term, $L(y,y_{target};\Theta)$ in the loss function seeking to find the loss minima, while the second term, $||\frac{\partial L(y,y_{target};\Theta)}{\partial \Theta}||_{2}$ reflects the sensitivity of the objective function to changes in the parameters. This term aims to identify parameters that lie in the neighborhoods with uniformly low loss values. As shown in Figure.\ref{fig1}d, this additional term $\frac{\partial L(y,y_{target};\Theta)}{\partial \Theta}$ can be expressed as $\frac{\partial L(y,y_{target};\Theta)}{\partial W} \cdot \frac{\partial W}{\partial \Theta}$, where $\frac{\partial W}{\partial \Theta}$ can optimize the stability of individual components and $\frac{\partial L(y,y_{target};\Theta)}{\partial \Theta}$ optimizes the stability of the whole system. Here the component stability refers to how the weight of a single component remains stable under small control parameter perturbations, while the system stability reflects how the final loss remains stable under small weight perturbations. Both types of stability are improved during optimization.

However, directly calculating the second term requires computing second-order derivatives, the Hessian matrix, which leads the computational complexity to increase from $O(n)$ to $O(n^2)$, where $n$ denotes the matrix size. This additional cost makes Equation.\ref{eq2} challenging to implement in practice. To address this, Foret et al. propose approximating the gradient for minimizing the loss function using a first-order gradient of $L$ at $\Theta+\Delta \Theta$~\cite{foret2021iclr}, the gradient is calculated through Equation.\ref{eq3} and Equation.\ref{eq4},

\begin{equation}
\Delta \Theta = r\frac{\partial L/\partial \Theta}{||\partial L/\partial \Theta||_2}
\label{eq3}
\end{equation}

\begin{equation}
\frac{\partial L_1}{\partial \Theta} = \alpha_1 \frac{\partial (\Theta + \Delta \Theta)}{\partial \Theta}|_{\Delta \Theta = r\frac{\partial L/\partial \Theta}{||\partial L/\partial \Theta||_2}}
\label{eq4}
\end{equation}
Here $\Delta \Theta$ represents the parameter perturbation that induces the largest change in the loss within the neighborhood of $\Theta$ and $r, \alpha_1$ are hyperparameters. $\partial L/\partial \Theta$ depicts the loss gradient with respect to parameter $\Theta$.

Minimizing the loss function can thus be achieved through a two-step process of automatic differentiation, as illustrated in Figure.\ref{fig1}e. The first step involves identifying the point $\Theta+\Delta \Theta$ that results in the maximum loss within the neighborhood (denoted as ‘Mid point’ in Figure.\ref{fig1}e) using Equation.\ref{eq3}. Once these parameters are identified, the gradient is recalculated at ‘Mid point’. This second gradient calculation with Equation.\ref{eq4} reflects the direction in which the maximum loss in the neighborhood can be reduced, i.e., directing to the 'flat minimum”, as shown at the point labeled ‘Low sharpness minimum’ in Figure.\ref{fig1}e. Consequently, this two-step BP not only finds the minimum loss but also ensures that the loss remains very low in the vicinity of the minimum, achieving a ‘flat minimum’. 

As shown in Figure.\ref{fig1}f, SAT is applicable to both mainstream offline and online training workflows. In offline training, the training process is conducted on the mathematical model $y=F(x;\Theta)$ of a PNN to directly determine the controlling parameters. Even when the mathematical model deviates from the actual system, the trained parameters remain valid and maintain high accuracy when applied to the physical system. In online training, the loss function is calculated directly from measurements, and the two-step differentiations are obtained using the gradient estimation methods mentioned earlier. We further demonstrate that, compared to conventional online training, SAT not only achieves the same accuracy on the specific physical device being trained but also enables the trained parameters to be transferrable to other devices. This contrasts with other online training methods.

\subsection{Resilient offline training with imperfect digital model}

We first demonstrate our approach’s effectiveness in the offline training scheme. In this scheme, where the mathematical model diverges from the actual system, directly implementing the offline trained model causes significant accuracy reduction using traditional methods. Moreover, any errors during parameter deployment and perturbations post deployment, such as thermal drift or alignment errors, will make the trained parameters inaccurate, leading to significant accuracy loss and requiring retraining. To show our approach’s effectiveness in addressing these problems, we use MRR-based PNNs as our first demonstration platform.

\begin{figure*}[ht]
\centering
\includegraphics[width=0.95\textwidth]{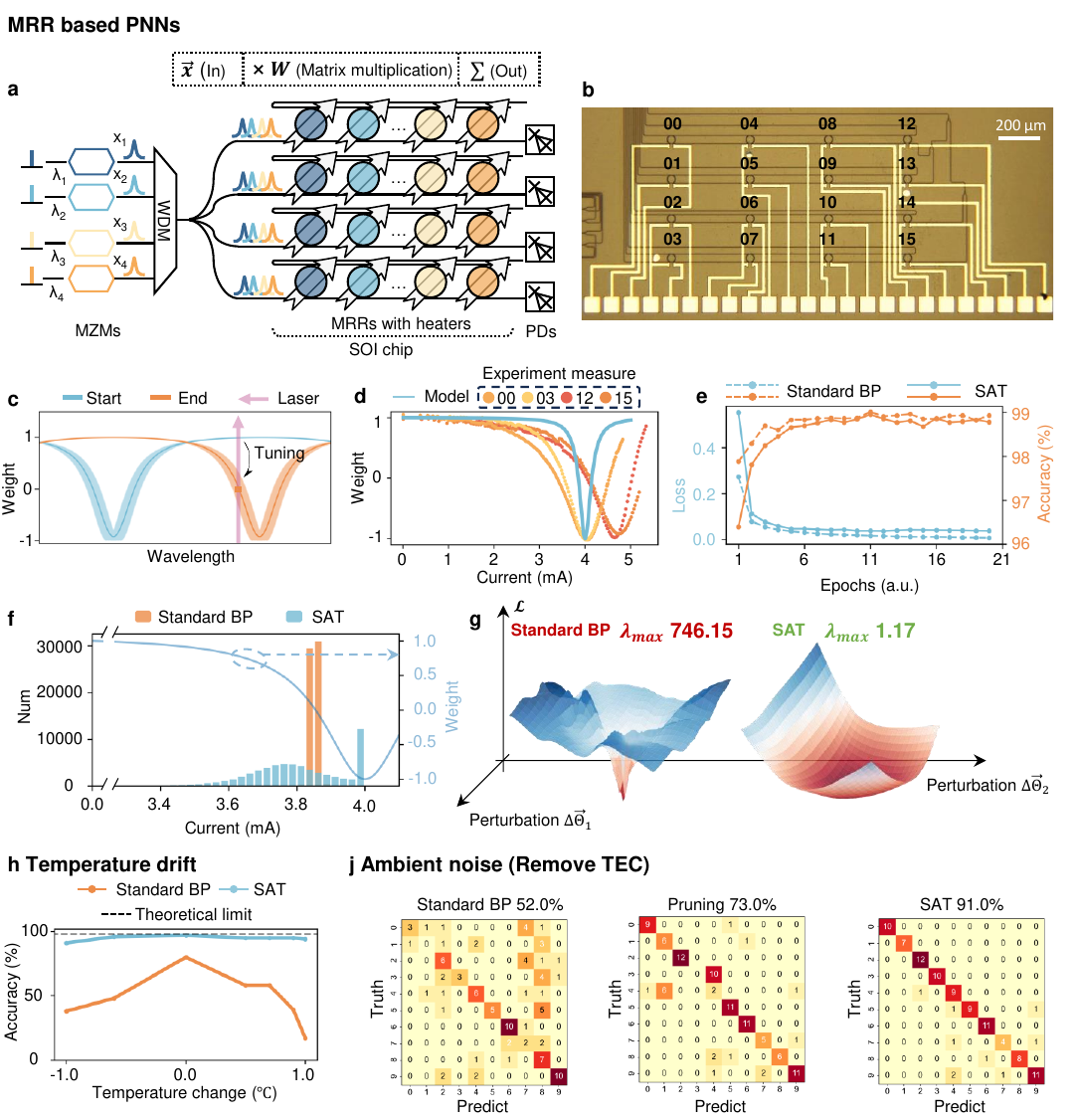}
\caption{Experimental results in MRR-based PNNs. (a) Implement matrix multiplication on MRR weight bank. (b) Picture of the manufactured MRR weight bank. (c) Single MRR spectrum changes through thermal tuning. (d) Theoretical MRR tuning curve model and experimental measured MRR tuning curves. (e) Loss and classification change verse training epochs. (f) Current distribution contrast histogram. (g) Trained model loss landscapes. (h) Experimental inference accuracy with temperature change from 21°C to 23°C. (j) Experimental measured inference accuracy without using TEC. MZM: Mach-Zehnder modulator. MRR: Micro-ring resonator. SOI: Silicon on insulator. PD: Photodetector. BP: Backpropagation. SAT: Sharpness-Aware Training. TEC: Thermoelectric control.}
\label{fig2}
\end{figure*}

In the MRR-based PNNs, weight synapses are realized using MRR weight banks, as illustrated in Figure.\ref{fig2}a. The input matrix is represented by an array of analog signals, each modulated onto a laser at a distinct wavelength. The multiplication operation is performed by an array of MRR weight banks, shown in Figure.\ref{fig2}b. Each MRR in the weight bank has a slightly offset radius, resulting in a unique resonance wavelength that aligns with the corresponding input laser's wavelength. By tuning the resonance wavelength of the MRR, the distribution of light between the Drop and Through ports can be precisely controlled, allowing for tunable weights, as shown in Figure.\ref{fig2}c. Thermal tuning is the most commonly employed mechanism for controlling MRRs due to its ease of implementation on silicon photonics chips. Once the input signals are weighted, they are detected by a balanced photodetector~\cite{huang2021ne,xu2024optica}. In this experiment, we use singal-end photodetector to detect the light from the Through port. MRR-based PNNs offer several advantages, including power-efficient tuning, straightforward weight assignment, and high computational density due to the compact device footprint~\cite{zhou2022light}. However, MRRs are highly sensitive to thermal changes, such as thermal drift, thermal crosstalk, and ambient disturbances. In our previous work~\cite{xu2024optica}, we observed that a resonance shift of just 12 nm (corresponding to a temperature change of 0.2°C) caused the PNN's accuracy to drop from a theoretical 99.0\% to 67.0\% when classifying Modified National Institute of Standards and Technology (MNIST) dataset~\cite{lecun1998ieee}.

Here, we conduct our experiment using a 4$\times$4 MRR weight bank, as shown in Figure.\ref{fig2}b (refer to the Supplementary Note 4 for detailed device design and experimental process). To demonstrate the effectiveness of SAT to address challenging problems, we have several experimental settings. (1) When constructing the mathematical model $y=F(x;\Theta)$, we treat all MRRs as identical. In this model, $\Theta$ is an array of currents used to control each MRR. We experimentally measure the current-weight relationship on only one specific MRR and apply this relationship uniformly across all MRRs in the model. However, as shown in Figure.\ref{fig2}d, fabrication variations lead to different current-weight characteristics for each MRR, resulting in modeling errors. (2) We ignore all side effects that could introduce modeling errors. These side effects include thermal crosstalk between devices, noise, and imperfections in peripheral circuits such as wavelength-dependent gain in erbium-doped fiber amplifier and non-ideal modulations. Ignoring these factors further causes the mathematical model to deviate from the actual system. (3) After training, the deployed system experiences thermal drift, further impacting the system's performance.

We first compare the training performances of our SAT method with traditional training method using standard BP on the MNIST database. As shown in Figure.\ref{fig2}e, both methods converge after 20 training epochs with a training accuracy of 99.0\%. (see Supplementary Note 4 for training details). The current distributions obtained using the two methods are shown in Figure.\ref{fig2}f. The effectiveness of SAT is highlighted by the loss landscape illustrated in Figure.\ref{fig2}g. The loss landscape clearly shows that the system trained with SAT exhibits a significantly flatter profile, indicating enhanced stability and robustness against variations in control parameters compared to traditional training methods~\cite{yao2020ieee}. To emphasize the largest curvatures, the x and y axes represent the two principal directions of the Hessian matrix of the loss function with respect to the control currents, while the z axis corresponds to the loss function. This robustness is also reflected in the maximum eigenvalue of the Hessian matrix, $\lambda_{max}$. A smaller eigenvalue suggests a flatter loss landscape. With SAT, $\lambda_{max}$ reduces dramatically from 746.8 with standard BP to just 1.17, highlighting the effectiveness of SAT in identifying stable regions within the physical system.

After training, we deploy the trained currents to the MRR-based PNN to evaluate the inference accuracy. As mentioned in the experimental setup, there is a significant deviation between the model used for training and the actual system due to static errors and dynamic noise. Consequently, when using standard BP, the accuracy for MNIST classification drops to 80.0\% from the theoretical accuracy 98.0\%. In comparison, the accuracy achieves with SAT remains at 97.0\%, as shown in Figure.\ref{fig2}h. This result is obtained at a temperature of 22°C, the same temperature used to characterize the MRR current-weight relationship. To further simulate the real-world conditions under temperature drift, we vary the chip temperature using a thermoelectric controller (TEC) from 21°C to 23°C and test the inference accuracies, with results shown in Figure.\ref{fig2}h. The system trained with SAT demonstrates remarkable robustness to temperature changes, maintaining its computational accuracy, while the accuracy of the system trained with standard BP degrades significantly from 80.0\% to 7.0\% as the temperature varies from 21°C to 23°C.

Next, we directly remove the TEC and measure the accuracy under the room temperature of 20°C. Note that the training model is still characterized at 22°C with TEC. In this circumstance, despite the temperature change being as large as 2°C, SAT still maintains high accuracy at 91.0\%, in significant contrast compared to 52.0\% for standard BP. We also compare the SAT method with our previous work on optical pruning~\cite{xu2024optica}. The core idea in that work is to train the parameters of the MRR-based PNN to enhance the noise robustness of each MRR, which focuses solely on improving the stability of individual devices. Our simulations show that under large model deviations and temperature fluctuations, optical pruning achieves an accuracy of only 73.0\%, although this is still higher than standard BP. In contrast, SAT not only improves the stability of individual components but also enhances the overall system's stability by ensuring that the entire system’s loss function remains minimized in response to weight perturbations. Consequently, SAT achieves the highest accuracy at 91.0\%.

The results demonstrate two significant advantages of SAT compared to traditional training techniques. Firstly, our approach does not require highly accurate digital modeling of the physical system or considering side effects or system imperfection, yet it still outperforms the method requiring them. Secondly, the PNN trained by SAT is significantly more robust against dynamic noise, particularly temperature drift, during deployment.

\subsection{Extending SAT to PNNs without explicitly known models}

\begin{figure*}[ht]
\centering
\includegraphics[width=0.95\textwidth]{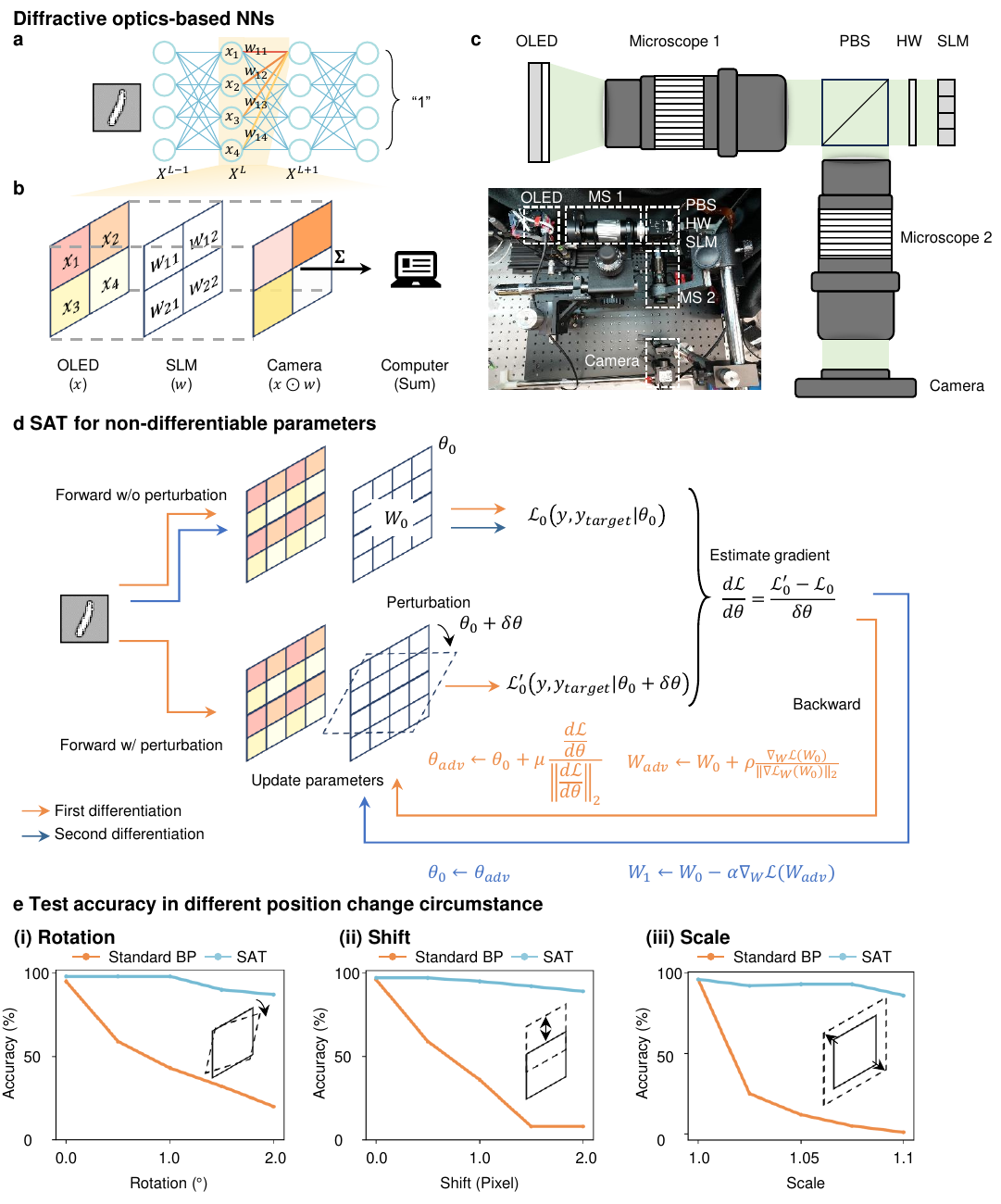}
\caption{Experimental results in diffractive optics-based NNs. (a) Decompose neural network inference into multiple matrix multiplication. (b) Deploy matrix multiplication in free-space optical computing systems. (c) Schematic diagram and picture of the experimental setup. (d) Detailed SAT optimization process for non-differentiable parameters. (e-i) Experimental inference accuracy with rotation angle change from 0.0° to 2.0°. (e-ii) Experimental inference accuracy with shift pixel number change from 0.0 to 2.0. (e-iii) Experimental inference accuracy with scaling number change from 1.0 to 1.1. OLED: Organic light-emitting diode. SLM: Spatial light modulator. MS: Microscope. PBS: Polarization beam splitter. HW: Half-wave plate. SAT: Sharpness-Aware Training. BP: Backpropagation. }
\label{fig3}
\end{figure*}

We further demonstrate our method’s effectiveness in diffractive optical NNs~\cite{wang2022nc}. In diffractive optical NNs, as shown in Figure.\ref{fig3}a and Figure.\ref{fig3}b, a NN layer is decomposed into multiple matrix multiplications and realized by a imaging system. In the experiment setup depicted in Figure.\ref{fig3}c, element-wise multiplication is achieved by first encoding the image to be processed on an organic light-emitting diode (OLED) serving as the light source. The optical image then passes through a spatial light modulator (SLM) which performs pixel-wise intensity modulation on the input image, thus realizing element-wise multiplication~\cite{wang2022nc}. (Details are explained in the Supplementary Note 5). In this setup, any misalignment between the OLED and SLM can severely degrade system performance. Our experiment shows that 1° rotation angle misalignment between the OLED and the SLM can cause a significant drop in accuracy, from 98.0\% to 43.0\% on MNIST dataset. Other misalignments include x and y axis shifts between the OLED and the SLM and the distance change between them (i.e., z axis shift) that results in image scaling. 

Unlike the MRR-based PNN system, where the relationship between the weight and the applied current is known (though not entirely precise), in diffractive optical NNs, the relationship between the misalignment factors such as rotation angle and the weights is not straightforward. This makes the implementation of SAT in such a system more challenging. Mathematically speaking, characterizing the model $y=F(x;\Theta)$ is difficult, where $\Theta$ represents the misalignment factors such as rotation angle and axis shifts between the OLED and the SLM. To address this, we propose a modified SAT method in which the gradient of the loss with respect to the misalignment factor is estimated by the finite difference method. The optimization flow is shown in Figure.\ref{fig3}d using rotation angle as an example. During the first automatic differentiation process as noted by the yellow arrows, to estimate the gradient, a small perturbation $\delta \theta$ is added to the rotation angle, and the loss is calculated twice, once without this small perturbation and once with it. The gradient with respect to the rotation angle is then estimated using the finite difference method, as depicted in Figure.\ref{fig3}d. After obtaining this gradient $dL⁄d\theta$, the rotation angle and weights are updated according to Equation.\ref{eq5} and Equation.\ref{eq6},

\begin{equation}
\theta_{adv} = \theta_0 + \mu \frac{d L/d \Theta}{||d L/d \Theta||_2}
\label{eq5}
\end{equation}

\begin{equation}
W_{adv} = W_0 + \rho \frac{\bigtriangledown_{W}  L(W_0)}{||\bigtriangledown_{W}  L(W_0)||_2}
\label{eq6}
\end{equation}

Here $\theta_0$ and $W_0$ denote the initialized rotation angle and weights, $\theta_{adv}$ and $W_{adv}$ represent the adversarial rotation angle and weights that reflect the maximum loss within the initial parameters' neighborhood. $\mu$ and $\rho$ are scaling hyperparameters.

Next, we maintain the rotation angle at $\theta_{adv}$ during the second automatic differentiation process and conduct forward propagation, as noted by the blue arrows. Then the gradient with respect to weights is calculated again at $W_{adv}$. This calculated gradient is used to update the initial weights $W_0$. Finally, the rotation angle is reset to the initial value $\theta_0$ as the rotation angle remains constant once calibrated in the real-world system. The training is performed on a computer without considering side effects like dark noises. (The detailed training process is illustrated in Supplementary Note 5).

We apply the same optimization approach to address the issues of shift and scaling. Afterward, we deploy the trained model in our experimental setup to perform NN inference. During the experiment, we use the affine function to quantitatively adjust rotation, shift, and scaling parameters. The experimental results shown in Figure.\ref{fig3}e illustrate that SAT significantly improves the model’s robustness against the misalignment issues in the free-space computing system.   Specifically, SAT maintains MNIST classification accuracy at 98.0\% at 1° rotation angle misalignment without any accuracy reduction, compared to 43.0\% using the standard training method. In terms of pixel-wise shift, we intentionally shift the pixels of the SLM from 0 to 2 pixels in the y-axis direction. The results show SAT achieves 97.0\% accuracy at 1-pixel shift, while the standard training method attains only 36.0\% accuracy. Finally, we scale the image size in the SLM to emulate the z axis shift, with the scaling factor sweeping from 1.0 to 1.1. Our experimental results show that the standard training method achieves only 12.0\% accuracy, but SAT improves the accuracy to 93.0\% with the scaling factor at the level of 1.05. These results demonstrate SAT generally applicable to different computing systems, even for those systems without having an explicitly known model. The modified SAT can successfully address this issue by using the finite difference method to estimate the gradient of the loss with respect to those parameters.

\clearpage

\subsection{Facilitate transferable online training}

\begin{figure*}[ht]
\centering
\includegraphics[width=0.95\textwidth]{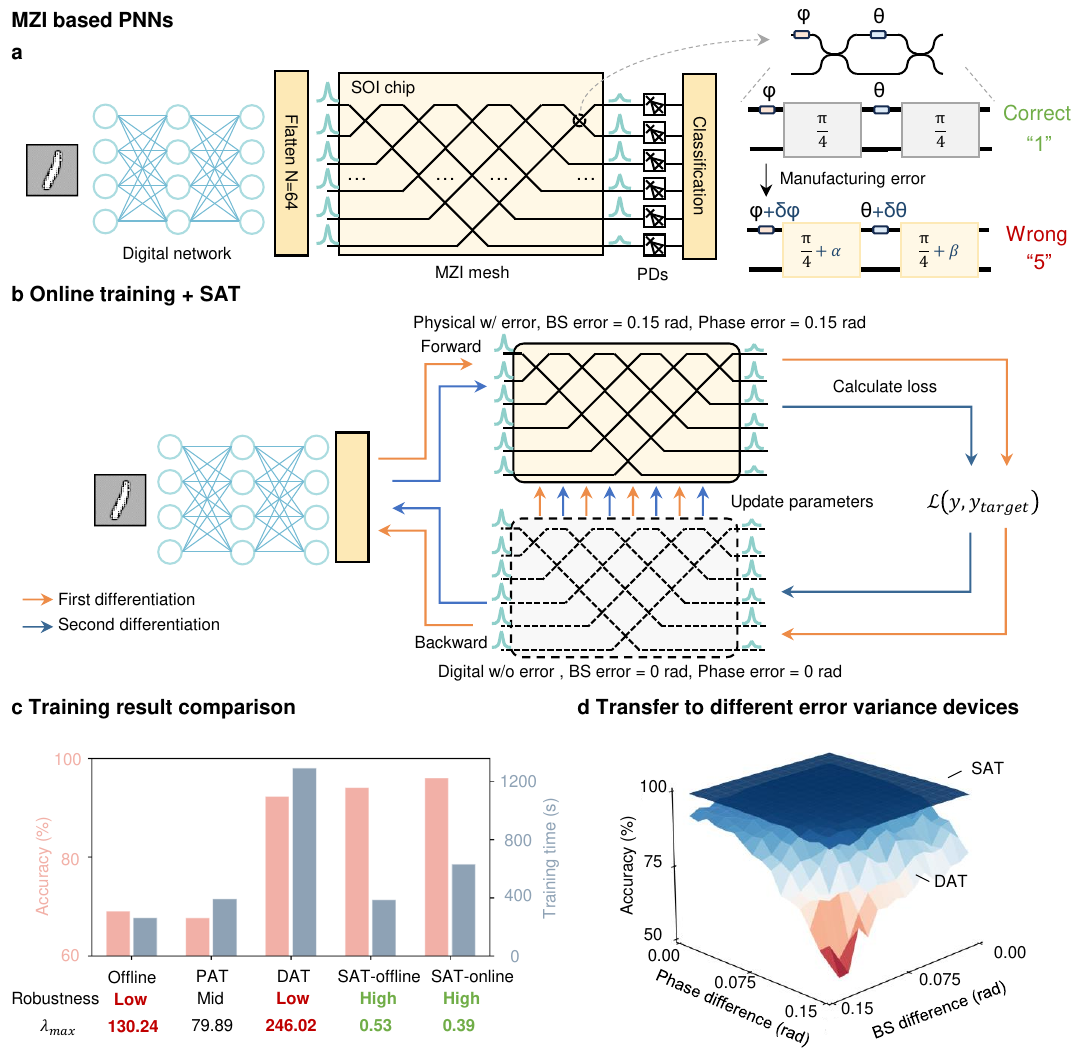}
\caption{Simulation results in MZI-based PNNs. (a) Schematic diagram of digital-optical hybrid network and illustration of manufacturing error. (b) Schematic diagram for implementing SAT on online training. (c) Training result performance comparison. (d) Inference performance comparison when transferring trained parameters to different error variance devices. MZI: Mach-Zehnder interferometer. PNN: Photonic neural network. SOI: Silicon-on-insulator. PD: Photodetector. SAT: Sharpness-Aware training. PAT: Physical-aware training. DAT: Dual-adaptive training. BS: Beam splitter.}
\label{fig4}
\end{figure*}

This section addresses a key challenge of online learning: the trained results are specific to a particular system and cannot be easily transferred to other systems, even those fabricated with the same design due to fabrication variances. This lack of transferability is a significant drawback for modern AI hardware, where training is typically performed on a single high-end computer or cloud-based cluster, and the trained parameters are then deployed across multiple edge devices for inference. Here we demonstrate that SAT can be combined with online training approaches to overcome this major drawback and enable transferable online training.

We demonstrate SAT on the MZI meshes based NN through numerical simulations using the library provided by Ziyang et.al.~\cite{zheng2023nmi}. In MZI meshes-based NNs, the basic building block is a 2$\times$2 MZI, which consists of two 50-50 beam splitters and two phase shifters. This block performs arbitrary 2$\times$2 unitary transformation by configuring two phase shifters. By cascading multiple MZI blocks, the MZI mesh can achieve arbitrary unitary transformations~\cite{clements2016optica}. Programming such a system remains challenging, as the model is difficult to characterize due to unknown initial phases caused by phase errors in the phase shifters. Consequently, many online training approaches have been proposed to program such systems. However, wafer-level manufacturing errors make the trained results are specific to only the trained NN but not transferrable to other systems. For example, as pointed out by~\cite{bandyopadhyay2021optica}, variations of beam splitter as small as 2\%, which is a typical wafer-level variance, can degrade the NN accuracy by nearly 50\%.  

We demonstrate our approach on simulated digital-optical hybrid NN, following the setup in~\cite{zheng2023nmi}. The digital NN consists of two convolutional layers followed by two fully connected layers, with the functions of feature extraction and image downsampling. The optical NN is a 64$\times$64 MZI mesh followed by a square detection layer, serving as the output layer. The first 10 output ports are used to generate classification output on the MNIST dataset. 

We evaluate five approaches under fabrication variances: (1) offline training with standard BP on a digital model without considering fabrication variances; (2) our SAT, also trained offline on a digital model without considering fabrication variances; (3) physical-aware training (PAT), an online approach that leverages the actual PNN system for forward propagation and a proxy model for gradient estimation; (4) dual-adaptive training (DAT), an online approach modified from PAT, capable of deriving a more accurate proxy model for improved gradient estimation; and (5) an online SAT, which adopts the PAT framework with additional training to also search for the sharpness minimum. Detailed setups for these approaches are clarified in Supplementary Note 6. 

To simulate the fabrication variances, we generate phase error $\sigma_{ps}$ and splitting error $\sigma_{bs}$ sampling from a Gaussian distribution with a mean of 0 and variance of 0.15, and add these errors to the MZI mesh model. As shown in Figure.\ref{fig4}c, when using offline training with standard BP, these fabrication errors cause a significant accuracy reduction from 97.4\% to 69.1\% This is because the fabrication variances create a substantial mismatch between the ideal model used for training and the actual system. The training time is 292.4 s. To address the model mismatch, we turn to online training methods including PAT and DAT, which loop the device in the training process. In PAT, the forward pass use the system with fabrication errors while the backward model is an ideal MZI mesh without error. Despite using online training, under large fabrication errors, PAT achieves an accuracy of only 67.7\% with a training time of 394.7 s. The low accuracy is due to the approximated gradients from the ideal model significantly deviating from the true gradients, which hinders the loss minimization. To obtain a more accurate proxy model for BP, DAT combines the ideal model with an additional digital model using the data driven method. DAT brings the approximated gradients closer to the real gradients, therefore improving the accuracy to 92.3\%. But this improvement is at the cost of adding six additional systematic error prediction networks (SEPNs) with the training time increasing to 1295.1s. The results are consistent with~\cite{zheng2023nmi}. We also find the model trained by DAT is highly sensitive to noises and perturbations. We use the maximum eigenvalue of the Hessian matrix $\lambda_{max}$ to evaluate the trained model, as a large $\lambda_{max}$ indicates high sensitivity to parameter changes. The $\lambda_{max}$ for DAT is 246.02, nearly twice that of the model trained offline with standard BP which has a $\lambda_{max}$ of 130.24. This result implies that while data-driven methods provide accurate gradients and improve classification performance, this comes at the cost of reduced robustness. As a result, the trained system will be highly vulnerable to perturbations and noise after deployment and the trained parameters cannot be transferred to other systems, even those fabricated with the same design. 

Next, we evaluate our SAT trained offline on an ideal model without considering fabrication variances, as well as online SAT adopting the PAT framework but with additional training to search for the sharpness minimum. The purpose is to verify: (1) whether SAT with offline training can overcome the fabrication errors and even outperform current online training methods; (2) whether SAT can integrate with existing online training methods, particularly PAT, and improve their performance in light of inaccurate modeling; (3) whether the online SAT can solve the major drawback of current online training methods of lack of transferability to other devices.

We first implement SAT trained offline on an ideal model. The training time is 386.7s, which is longer than that of offline training with standard BP due to the need for two steps of BP. However, the trained system is much more robust to the fabrication errors, achieving an accuracy of 94.1\% even under large fabrication errors. Remarkably, this performance not only outperforms the offline training methods but also surpasses online training methods including PAT and DAT. Moreover, the Hessian max eigenvalue $\lambda_{max}$ is only 0.63, which is significantly lower than other methods under comparison. The result shows the trained model has high robustness compared to others.

To further improve accuracy, we combine SAT with PAT, resulting in an accuracy of 96.1\%. This outcome suggests that incorporating SAT effectively addresses the issue in PAT, where the proxy model used for BP deviates significantly from the actual system. Moreover, SAT demonstrates the ability to achieve high training performance even with inaccurate approximating gradients. Moreover, the Hessian max eigenvalue $\lambda_{max}$ is only 0.39, indicating that SAT significantly improves the robustness of the trained system. 

This robustness allows the parameters obtained through online training to be transferred to other devices without a loss in accuracy. To evaluate this, we apply the trained model to other devices with different randomly generated phase errors and splitting ratio errors. We systematically increase the errors in both phases and splitting ratio and assess classification accuracy, with results shown in Figure.\ref{fig4}d. SAT is capable of maintaining high accuracy at 95.2\%, indicating that SAT can facilitate transferable online training. In contrast, the system trained by DAT decreases to 58.6\% when the phase and splitting ratio errors reach 0.15.

Overall, SAT significantly enhances both offline and online training schemes. When trained offline, SAT achieves high accuracy under large fabrication errors, surpassing state-of-the-art online methods. Moreover, integrating SAT with online training further improves accuracy. Importantly, SAT enables the trained parameters to be transferable to other devices, addressing a key limitation of current online training methods.

\section{Discussion and conclusion}\label{sec3}

In summary, we propose a novel training method that addresses key limitations in both offline and online training of PNNs. Traditional offline training requires highly accurate digital modeling of physical systems, yet the inference accuracy often falls significantly below the theoretical value when the model is deployed. Online training improves accuracy by integrating the specific hardware into the training loop. However, online training require extensive measurement or additional training costs for gradient estimation. Moreover, we find that the models trained online are highly sensitive to perturbations and noise, making it difficult to transfer the trained model to other devices, even with identical design. This is problematic in modern AI hardware landscape, where training is typically conducted on a high-end computer or cloud-based computer clusters, and the trained parameters are then deployed across many edge devices for inference. Additionally, both offline and online training face the issue that deployed systems are sensitive to perturbations such as thermal drift or alignment errors, requiring retraining to recover accuracy. 

The proposed SAT method addresses all these challenges. Through demonstrations on three typical PNNs, an MRR-based PNN, a diffractive optical NN, and an MZI mesh-based PNN, we highlight several distinguishing features of SAT that offer clear advantages over current offline and online training methods: (1) SAT enables accurate offline training of a PNN, even with an imprecise digital model and under deployment errors; (2) SAT allows systems trained offline to outperform those trained online, even in the presence of significant fabrication errors; (3) systems trained using SAT exhibit high robustness to environmental fluctuations and noise; (4) SAT can be integrated with online training methods, facilitating transferable online training across different devices; and (5) SAT is universally applicable to various PNNs, regardless of whether their models are explicitly known or unknown. Overall, SAT offers a practical, effective, and computationally efficient solution for training and deploying PNNs in real-world applications.

While SAT has a slight cost in additional automatic differentiation during training, it remains far more efficient than methods requiring extensive experimental measurements or the training of a large digital twin model. However the additional training steps can be further reduced using the approach proposed in~\cite{du2022nips}, which reduces the sharpness with almost zero additional computational cost by calculating the Kullback–Leibler-divergence between the neural network’s outputs using current weights and past weights, as a substitute for SAT’s sharpness measure. This approach will offer the benefits of reduced sharpness without extra overhead.

\section{Methods}\label{sec4}

\subsection{Device fabrication}

The device is fabricated in the commercial multi-project wafer (MPW) process from the Applied Nanotools (ANT). The silicon thickness of this process is 220 nm, and the buffer oxide layer thickness is 2 µm. The MRR weight bank contains 16 MRRs, and is arranged in a 4×4 matrix. Each row sub-weight bank consists of four MRRs coupled with two bus waveguides in an add/drop configuration. The radii for each MRR is around 20 µm, and we intentionally introduce a slight 10 nm radius difference to avoid resonance collision. And the result shows resonance wavelengths are roughly spaced by 0.8 nm. The gap between the MRR and the bus waveguide is 100 nm. MRR Q factor is around $10^4$. Circular metal heaters (TiW) are built on top of each MRR for thermal weight tuning. Metal (Al) vias and traces are deposited to connect heater contacts of the MRR weight bank to electrical metal pads. The heater resistance is around 375 Ohms. The tuning efficiency is around 0.53nm/mW.

\subsection{MRR-based PNN experimental setup}

During the experiment, we select 4 MRRs with the labels ‘00’, ‘03’, ‘12’, ‘15’. Before the experiment, we first actuate initialization currents to align the resonance wavelengths of ‘00’ $\&$ ‘03’ and ‘12’ $\&$ ‘15’. After initialization, the resonance wavelengths of ‘00’ $\&$ ‘03’ are both 1542.7nm, and the resonance wavelengths of ‘12’ $\&$ ‘15’ are both 1544.8nm. A TEC module is set to stabilize or intentionally change the temperature. The laser (KG-TLS-C-13-50-P-FA, Conquer) wavelengths are set to 1543.0nm and 1545.0nm, respectively. Different wavelength lights are independently modulated using electrical-optical modulation (KG-DDMZM-RF-B, Conquer). The modulation signals are set with a baud rate of 20 Baud (AFG 31000 SERIES, Tektronix). The different wavelength lights are then combined through a 50:50 coupler. Before coupling into the chip, the combined light is amplified through an EDFA (FA-30, PriTel) to around 12dBm. The coupling loss of the chip is around 14dBm. The two-channel output lights are then detected by the photodetector (PDT0313-FC-A, HP) at the through port. During the matrix multiplication, the negative weights are mapped to the tuning range [0,1].

\subsection{Free-space experimental setup}

In the free-space experimental setup, a custom OLED display (Sony ECX335S) is used, encoding pictures as the light source. Only the green channel is utilized in the experiments. The display has a resolution of 1920 $\times$ 1080 and a refresh rate of 60Hz. We develope custom Python control codes to load images. The display features 256 distinguishable brightness levels, corresponding to an 8-bit resolution. We achieve the weights through light intensity modulation using an SLM (LCOS-SLM, HDSLM80R Plus, UPO labs) with a polarizing beam splitter (PBS) and a half-wave plate (HW) in a double-pass configuration. A zoom-lens microscope (BYH0330, Inseinlifung) is used between the OLED and SLM to match the pixel sizes between the OLED (5 $\mu$m) and SLM (8 $\mu$m). The light is then projected onto a camera using a second microscope (SHL-0745C, Shhunhuali). A CMOS camera (MV-SUF89OGC/M, Mindvision) captures the modulated light field as an image. The image is separated into different regions, and the sum of the intensities from these regions represent the matrix multiplication result. 

\subsection{Details about training the digital-optical hybrid network}

The selected task is MNIST, which consists of 70,000 grayscale 28$\times$28 pixel images, with 60,000 images used for training and 10,000 for testing. The digital-optical hybrid network is built based on the open-source library 'DAT MPNN'~\cite{zheng2023nmi} and 'Neurophox'~\cite{pai2019pra}. The network is a modified version of the classic LeNet-5 architecture, designed for image classification tasks, with an optical MZI mesh layer for classification. Details are clarified in Supplementary Note 6.

Specifically, we test 5 different training methods: offline standard BP, PAT, DAT, offline SAT, and online SAT. The training device is a Nvidia 3070Ti, and the training time is measured by directly calculating the difference between the start time and end times of the program run. The offline standard BP involves training the ideal model directly for 30 epochs. For PAT, training involves performing inference on the physical system and BP on the ideal system. The physical system deviates from the ideal system due to static phase errors and splitting errors. The total number of training epochs is 40 for accuracy and loss to converge. 

In the case of DAT, an additional digital network SEPN is introduced to compensate for the gap between the physical system and the ideal system. DAT requires more training epochs to converge, totaling 60 epochs. During the first 30 epochs, the training process is the same as the PAT, and SEPN is trained to reduce the output difference between the physical system and the ideal system, but it is not used for gradient calculations. In the subsequent 30 training epochs, SEPN training stops, and SEPN assists the ideal model in providing accurate gradients. To minimize the gap as much as possible, we use 6 complex mini-UNet networks connected in parallel to extract multiscale features. The detailed network of mini-UNet is the same as that in~\cite{zheng2023nmi} and is illustrated in Supplementary Note 6.

For offline SAT, standard BP is still used for differentiation, but differentiation is performed twice per epoch. The total number of training epochs remains 30. For online SAT, differentiation is provided by the ideal model, this is the same as PAT, and differentiation is performed twice per epoch. The total number of training epochs is 40. The initial learning rate for training the PNN is 0.1, which is decayed by 0.5 at half of the maximum training epoch number. The learning rate for training the SEPNs is set at 0.001 for constant. The cross-entropy loss function is selected to calculate the task loss, and the MSE loss function is selected for adaptive training of the SEPNs.

\section*{Acknowledgements}

This work was supported RGC ECS 24203724, NSFC 62405258, ITF ITS/237/22, ITS/226/21FP, RGC YCRF C1002-22Y, RNE-p4-22 of the Shun Hing Institute of Advanced Engineering, NSFC/RGC Joint Research Scheme N CUHK444/22 and CUHK Direct Grant 170257018, 4055143.

\section*{Author contributions}

Z.L., T.X. and C.H. conceived the ideas. T.X. and S.L. built the MRR-based PNN computing system with the help of Q.X. and B.W.. Z.L. and L.F. built the diffractive optics-based NN computing system. T.X., S.L., Z.L., and C.H. designed and conducted the experiments of the MRR-based PNN. Z.L., L.F., T.X., and C.H. designed and conducted the experiments of the diffractive optics-based NN. T.X., Z.L., S.L., L.F., Q.X. and B.W. analyzed the results. D.W. designed the PCB board and performed the chip packaging. T.X., S.L. and C.H. wrote the manuscript with input from all authors.  C.H. supervised the research and contributed to the general concept and interpretation of the results. All the authors discussed the data and contributed to the manuscript.

\section*{Data availability}

The data and code that support this study are available from the corresponding authors upon reasonable request.

\section*{Conflict of interest}

The authors declare no conflict of interest.

\bibliography{sn-bibliography}

\end{document}